
\input epsf


\input mn


\let\sec=\section
\let\ssec=\subsection


\def\secno#1{\vskip 1cm\noindent{\bf #1}\vglue 0.3cm\noindent}
\def\bigstrut{\vrule width0pt height0.35truecm}
\font\japit = cmti10 at 11truept
\def\ss{\scriptscriptstyle\rm}
\def\ref{\parskip =0pt\par\noindent\hangindent\parindent
    \parskip =2ex plus .5ex minus .1ex}
\def\gs{\mathrel{\lower0.6ex\hbox{$\buildrel {\textstyle >}
 \over {\scriptstyle \sim}$}}}
\def\ls{\mathrel{\lower0.6ex\hbox{$\buildrel {\textstyle <}
 \over {\scriptstyle \sim}$}}}
\newcount\equationo
\equationo = 0
\def\leftdisplay#1$${\leftline{$\displaystyle{#1}$
  \global\advance\equationo by1\hfill (\the\equationo )}$$}
\everydisplay{\leftdisplay}

\def\kmsmpc{\;{\rm km\,s^{-1}\,Mpc^{-1}}}
\def\hompc{\,h\,{\rm Mpc}^{-1}}
\def\mpcoh{\,h^{-1}\,{\rm Mpc}}

 

\def\annrev{ARA\&A}

\def\apj{ApJ}

\def\mn{MNRAS}

%

\pageoffset{-0.8cm}{0.2cm}




\begintopmatter  

\vglue-2.2truecm
\centerline{\japit Accepted for publication in Monthly Notices of the R.A.S.}
\vglue 1.7truecm

\title{Nonlinear evolution of cosmological power spectra}

\author{J.A. Peacock$^1$ and S.J. Dodds$^2$}

\affiliation{$^1$Royal Observatory, \bigstrut Blackford Hill, Edinburgh EH9 3HJ\hfill\break
$^2$Institute for Astronomy, University of Edinburgh,
Blackford Hill, Edinburgh EH9 3HJ\hfill\break
}

\shortauthor{J.A. Peacock and S.J. Dodds}

\shorttitle{Nonlinear evolution of cosmological power spectra}


\abstract{%
Hamilton et al. have suggested an invaluable scaling formula which describes
how the power spectra of density fluctuations evolve into
the nonlinear regime of hierarchical
clustering. This paper presents an extension of their method to
low-density universes and universes with nonzero cosmological
constant. We pay particular attention to models with large
negative spectral indices, and give a spectrum-dependent fitting
formula which is of  significantly improved accuracy by
comparison with an earlier version of this work.
The tendency of nonlinear effects to increase power on small
scales is stronger for spectra with more negative spectral
indices, and for lower densities. 
However, for low-density models with a cosmological constant, the
nonlinear effects are less strong than for an open universe
of the same $\Omega$.
}

\maketitle  

\sec{INTRODUCTION}

The power spectrum of density fluctuations is a statistic of
central importance in cosmology, as it describes a combination of
the primordial deviations from homogeneity and their subsequent
modification by the matter content of the universe. One
practical obstacle to a measurement of this useful function is that
the present universe occupies a state where nonlinear gravitational
growth of density fluctuations has altered the form of the
fluctuation spectrum.

Prior to 1991, it was assumed that this problem was tractable
only in two extreme limits. On large scales, linear theory
applies and the power spectrum scales as the square of the
density growth factor.
For very large wavenumbers, we enter the `stable clustering'
regime, where there is a simple relation between the
power-law index of the linear spectrum and that of the
output nonlinear spectrum (see Section 73 of Peebles 1980).
It was therefore a significant breakthrough when Hamilton
et al. (1991; HKLM) suggested a scaling procedure which
allowed an accurate description of the transition regime between
these two limits in terms of an empirical universal function.

In a previous paper (Peacock \& Dodds 1994; PD), we extended the
HKLM procedure in a number of ways. First, we presented a version
of the method which worked with power spectra, rather than HKLM's
choice of integrated correlation function. Second, we
considered the modifications to the HKLM argument needed to work
in a universe of arbitrary density, rather than HKLM's 
$\Omega=1$. We applied this extended HKLM method to
a compilation of clustering data, and concluded that the linearized data
for $k\ls 0.4 \hompc$ ($h\equiv H_0/100 \kmsmpc$) were consistent with a spectral index of
about $-1.5$ on small scales, steepening to close to the primordial
$n=1$ on large scales ($P(k)\propto k^n$).

However, the universal scaling procedure of HKLM was known to fail for more
negative spectral indices, as has been investigated in some detail recently
for $\Omega=1$ models by Jain, Mo \& White (1995; JMW). A common
description of the linear power spectrum is in terms of the CDM
model, whose power-law slope tends to $-3$ on very small scales, so 
the HKLM method and its extensions in PD may not work
very well if we attempt to investigate scales significantly smaller
than those studied in PD. We have therefore run an additional
ensemble of $N$-body simulations which concentrate on the case of
spectra with $n\ls -1$ and the CDM spectrum. It turns out that a
simple slope-dependent correction can be made to the PD formulae which
provides an excellent description of the nonlinear data over essentially
all regimes of interest, and this is described below.

\sec{THE HKLM METHOD}

The key argument of HKLM is that gravitational collapse
causes a change of scale. By regarding the integrated
correlation function $\bar\xi(r)$ as measuring the 
number of excess neighbours within radius $r$, they
suggested that observed nonlinear correlations on 
nonlinear scale $r_{\ss NL}$ be related to a
pre-collapse linear scale via
$$
r_{\ss L} =[1+\bar\xi_{\ss NL}(r_{\ss NL})]^{1/3} r_{\ss NL}.
$$
HKLM then conjectured that, having translated scale, the
linear and nonlinear correlations had some universal relation
$$
\bar\xi_{\ss NL} (r_{\ss NL}) = f_{\ss NL}[\bar\xi_{\ss L}(r_{\ss L})].
$$
The function  $f_{\ss NL}$ must behave as
$f_{\ss NL}(x)=x$ in the linear $x\ll 1$ limit and $f_{\ss NL}(x)\propto x^{3/2}$
in the stable-clustering $x\gg 1$ limit, and must be determined numerically
around $x\sim 1$.

PD argued that a very similar argument could be made to work
for power spectra, using a dimensionless version of
the power spectrum: $\Delta^2$ is the
contribution to the fractional density variance
per unit $\ln k$.  In the convention of
Peebles (1980), this is
$$
\Delta^2(k)\equiv{{\rm d}\sigma^2\over {\rm d}\ln k} ={V\over (2\pi)^3}
\, 4\pi \,k^3\, |\delta_k|^2
$$
($V$ being a normalization volume), and the relation to the correlation function is
$$
\xi(r)=\int \Delta^2\; {dk\over k}\; {\sin kr\over kr}.
$$
Since $\bar\xi(r)$ can be thought of as $\Delta^2$ at some
effective wavenumber, this suggests the $k$-space version of HKLM:
$$
k_{\ss L} = [1+\Delta^2_{\ss NL}(k_{\ss NL})]^{-1/3} k_{\ss NL},
$$
$$
\Delta^2_{\ss NL}(k_{\ss NL}) = f_{\ss NL}
[\Delta^2_{\ss L}(k_{\ss L})].
$$

The extension to models with $\Omega\ne 1$ is straightforward
in the highly nonlinear regime. The $f_{\ss NL}(x)\propto x^{3/2}$
scaling comes because, once a virialized object is formed, the
nonlinear correlations depend on scale factor $a(t)$ just as the
background density, $\bar\xi_{\ss NL}\propto a^3$, whereas the linear
correlations scale as $\bar\xi_{\ss L}\propto a^2$. If we now allow
a density-dependent growth-suppression factor into the linear
growth law, 
$$
\bar\xi_{\ss L}\propto [a\,g(\Omega)]^2,
$$
the virialized regime becomes
$$
f_{\ss NL}(x)\propto x^{3/2} [g(\Omega)]^{-3}.
$$
The $g(\Omega)$ factor may conveniently be taken from the
high accuracy fitting formula of
Carroll, Press \& Turner (1992):
$$
g(\Omega) ={{5}\over{2}}\Omega_{\rm m}\left[\Omega_{\rm m}^{4/7}-\Omega_{\rm v}+
(1+\Omega_{\rm m}/2)(1+\Omega_{\rm v}/70)\right]^{-1},
$$
where we have distinguished matter ($m$) and vacuum
($v$) contributions to the total density parameter.

The problem is now reduced to one of running a number of
$N$-body simulations to obtain the full form of $f_{\ss NL}$
and to investigate its dependence on cosmological model.
The power of the original HKLM method was the belief that
$f_{\ss NL}$ was a universal function. Using this assumption, PD 
fitted a $g$-dependent nonlinear function to  a restricted
set of $N$-body simulations -- a procedure which works reasonably
well for spectra with $n$ in the region of $-1$.
However, with the larger library of simulations studied here,
it is possible to see that there is a
dependence on the linear spectrum, but one that can be
described well by a simple fitting formula.

\sec{NUMERICAL DATA AND FITS}

\ssec{$N$-body code}

The carrying out of the required set of numerical experiments
has become much easier recently, partly due to increased power
of computer hardware, but mainly through the generous distribution
of the Adaptive Particle-Particle Particle Mesh (AP3M) code
of Couchman (1991). This solves the Poisson equation on a mesh
to obtain the large-scale force, which is then supplemented by
exact pairwise forces from the near neighbours. Such P3M codes
normally slow down for highly clustered distributions, but this
is avoided in Couchman's method by adaptive regridding of the
densest regions on a finer mesh.

For the present purposes, the main problem with
Couchman's code is that it is supplied for an $\Omega=1$
cosmology only. The equations of motion use positions, $\bf x$,
measured in units of the box size, $L_{\rm box}$ (plus a
scaling with the Fourier mesh size). The velocities are
defined with these length units and a time unit of the
age of the universe at the start of the simulation.
Also, a `time' variable, $p$, is used to integrate the equations of motion:
$$
p={3\over 2\alpha}\; a^\alpha.
$$
In principle, the index $\alpha$ can be altered according
to the spectrum; we retained $\alpha=3/2$, so that both $p$
and the scalefactor $a$ were unity at the start of the simulation.

The standard equation of motion (e.g. Section 14 of Peebles 1980)
can then with a little effort be cast into the form (dashes denoting $d/dp$)
$$
{\bf v}' +\left[{1\over p} + {2\over\alpha p} + {T'\over T}\right]\;{\bf v} = 
\Omega_m(p)\, \left({3\over 2\alpha p}\right)^2\, {\bf f},
$$
where $\bf f$ is the force vector as calculated by the AP3M program,
and $T=H(p)t_i$ is the product of the physical Hubble parameter and the initial time. 
The required modifications to the AP3M equations of motion are therefore
to multiply the forces on the rhs by $\Omega_m(p)$ and to use the
appropriate $T(p)$. Note that a non-zero cosmological constant only enters through $T(p)$.

For an Einstein-de Sitter Universe, $T=(2/3)a^{-3/2}$.  In general,
we have the exact result (Carroll et al. 1992)
$$
H(a)=H_0 \sqrt{\Omega_{\rm v}(1-a^{-2})+\Omega_{\rm m}(a^{-3}-a^{-2})+a^{-2}}.
$$
There is also the excellent approximation
$$
H_0 t_0={2\over 3}\; |1-f|^{-1/2}\; S_k \sqrt{|1-f|\over f},
$$
where $f=0.7\Omega_{\rm m}-0.3\Omega_{\rm v}+0.3$ and $S_k$ is sinh
if $f<1$, otherwise sin. However, this is not needed, since we are
only interested in $T'/T$:
$$
{T'\over T}= {1\over 2\alpha p}\;
{ \Omega_{\rm m}(2a^{-2}-3a^{-3})+2a^{-2}\Omega_{\rm v}-2a^{-2} \over
  \Omega_{\rm m}(a^{-3}-a^{-2})+\Omega_{\rm v}(1-a^{-2})+a^{-2} }.
$$

The only problem with using these formulae to obtain
$T(a)$ is Couchman's convention that the reference time $t_0$
at which $a=1$ is at the start of the simulation, so that
$\Omega_{\rm m}$ and $\Omega_{\rm v}$ in the above formulae are the
initial values, not those at the desired endpoint of the
simulation, designed to correspond to the present epoch. 
To relate the initial and final values of 
$\Omega_{\rm m}$ and $\Omega_{\rm v}$, we use
$$
\Omega_{\rm m}(a)= {\Omega_{\rm m} \over
a + \Omega_{\rm m}(1-a) + \Omega_{\rm v}(a^3-a) },
$$
$$
\Omega_{\rm v}(a)= {a^3 \Omega_{\rm v} \over
a + \Omega_{\rm m}(1-a) + \Omega_{\rm v}(a^3-a) }.
$$
These formulae work whether $a=1$ is regarded as the start or the
end of the simulation; the important thing is not to mix the conventions.

Lastly, there is the question of initial conditions. The
AP3M program generates a realization of a displacement field
corresponding to the given  power spectrum.
A possible source of confusion, as usual, lies in
power-spectrum units. Couchman's function {\tt pow} relates
to the notation here via
$$
{\tt pow}= {2\pi^2\over [L_{\rm box} k]^3}\; \Delta^2(k).
$$
Zeldovich initial conditions are assumed, so that inital displacement
and velocity are proportional. However, in order to be
properly in the growing mode, the $\Omega=1$ velocities
should be multiplied by the growth factor $\Omega_{\rm m}^{0.6}({\rm initial})$
(effectively independent of $\Omega_{\rm v}$: Lahav et al. 1991).

\ssec{Simulations and analysis}

There are many degrees of freedom in possible $N$-body
runs. The physical ones are the power spectrum of interest
and the cosmological model. The size of the simulation box
also matters, since this must be set so that the
fundamental mode does not saturate:
$$
\Delta^2(2\pi/L_{\rm box}) \ll 1.
$$
If this condition is violated, the results may not be
reliable on any scale, owing to the missing power beyond the
box scale; we used a maximum value of 0.04. 
It is also necessary that $\Delta^2$ on the initial mesh scale
does not exceed unity, so that the Zeldovich method used to
set up initial conditions does not produce excessive
shell crossing.
For steep spectra, this can imply a more restrictive
limit on the final box-scale amplitude.

There remain the numerical parameters of the number of
particles, the total expansion factor, the number of timesteps,
the force softening and the required force accuracy.
We carried out various experiments varying these parameters,
to see how robust the results were. We generally found
that even quite simple simulations would reproduce the main features
of the nonlinear results here, particularly the quasilinear regime.
We obtained most of our data in a standard configuration of
$N=80^3$ particles, integrated in 300 timesteps over an
expansion factor of 15. A $128^3$ Fourier mesh was used,
with the initial softening set at one cell, held constant
in proper terms down to a minimum of 0.1 cell. Some cases
were checked in $N=100^3$ runs with 600 timesteps, but these
much longer simulations gave essentially identical results.

The only factors which influenced the resulting
power spectra significantly were particle placement and
the question of whether to use proper Gaussian realizations
in the initial conditions. It is common to start $N$-body
simulations by applying the initial displacement field to
particles placed on a uniform grid, but there are times
when this is not desirable. For example, the original grid
is often noticeable even at late times in void regions.
A cosmetic improvement can be made by starting from a set
of particle positions which are irregular but sub-random,
such as the `glass' described by Baugh, Gazta\~naga \& Efstathiou (1995).
A simpler alternative is to displace each particle randomly
within its grid cell, which sets up an $n=2$ spectrum of
initial perturbations in addition to those imposed by the
initial displacement field. With any of these alternatives, some
care is needed in obtaining the power spectrum of the density
field. 

As in the case of redshift
 surveys,
the particle distribution is Fourier analyzed, and shot noise is
subtracted from the raw power to allow for particle discreteness
(see Peacock \& Nicholson 1991). 
This would be a correct procedure at all times
if the initial particle positions were
also Poisson distributed, but the resulting small-scale
fluctuations would then swamp the desired physical spectrum.
The alternative starting schemes cure this problem, but introduce
discreteness fluctuations that are initially smaller than Poisson,
so that subtracting shot noise underestimates the spectrum
in the early stages of the simulation.
Small-scale mixing rapidly cures this problem
at large $k$ (effectively as soon as the first pancakes form), but
the overall spectrum will not be correct until the large-scale
portion has grown enough to exceed the shot power
$$
\Delta^2_{\rm shot}={4\pi \over N} \;
[k/(2\pi/L_{\rm box})]^3.
$$
For $N=80^3$ particles, the shot power on the box scale is $\Delta^2_{\rm shot}=10^{-4.6}$,
and this only becomes negligible when the box-scale power is $\sim 10^{-3}$.
For $n<0$, the shot power rises more rapidly with $k$ than the physical
power, and so in practice larger values of the box-scale power need to
be used in order that the quasilinear part of the spectrum is not
affected by shot-noise subtraction. For $n=-2$, this requires box-scale
power at the nonlinear limit of 0.04: simulation of more negative indices
is not feasible without a great increase in the number of particles used.

Even for the analysis of $N\sim 10^6$ particles,
only an FFT is practical, in which case the range of $k$ accessible is
limited by the size of the FFT array. For a $256^3$
mesh (our normal limit), the Nyquist frequency is only 128 times the fundamental,
and even here the power spectrum is affected by binning and by
aliasing (Baugh \& Efstathiou 1994). The binning correction is,
to second order in $k$,
$$
\Delta^2_{\rm bin}=\Delta^2_{\rm true} / (1+k^2B^2/12),
$$
where $B$ is the bin size ($=$ simulation box / 256). 
Aliasing effects depend on the slope of the power spectrum;
gravitational instability tends to produce an effective index
in the range $-1$ to $-2$, so this is not a problem in practice.
Normally this size of FFT was sufficient for our purposes,
but occasionally we wished to extend the results to smaller scales.
We then divided the data into a set of sub-cubes and analyzed
each separately, obtaining an estimate of the small-scale
power spectrum by averaging. Although these samples are
individually quite nonlinear on their box scale, a few
comparisons with time-consuming direct Fourier transforms
of the complete dataset indicated that the overall power
spectrum could be recovered in this way over a factor of 300
in $k$, allowing the maximum range of nonlinearity
to be probed consistent with the resolution of the simulations.
It is interesting to note that high accuracy is required in
the Fourier analysis. A small error in nonlinear power 
produces a small error in the deduced linear power; however,
because $f_{\ss NL}$ is so steep in the quasilinear regime, 
this small horizontal shift can result in a gross vertical error in 
$f_{\ss NL}$. More comfortingly, this means that the forward
prediction of nonlinear power is quite robust with respect to
moderate errors in $f_{\ss NL}$, and our  simple parameterization
of the numerical results should give very accurate results in
most circumstances.

Finally, to create a proper realization of Gaussian initial perturbations,
each Fourier mode should have a random phase and a power exponentially
distributed about the mean. The alternative is to use exactly the
expectation power, which has the advantage that the large-scale
linear portion of the power spectrum does not suffer a large
scatter owing to the limited number of modes. The results on
intermediate and small scales seemed identical in either case, and
so we used `improper' realizations.

In this way, we built up a library of results covering
$-2<n<0$ and $0.1<\Omega_{\rm m}<1$, considering
both open models with $\Omega_{\rm v}=0$ and flat models with
$\Omega_{\rm v}+\Omega_{\rm m}=1$. For the same 
cosmologies, we also considered CDM spectra, which
are parameterized by the shape $\Omega h$ and the normalization
$\sigma_8$ (see PD). For a given cosmology, these are
degenerate degrees of freedom in that a reinterpretation of
the simulation box length will change both $\Omega h$ and $\sigma_8$.
It therefore suffices to fix one parameter and vary the
other. We considered what is roughly the observed shape,
$\Omega h=0.2$, in a $100 \mpcoh$ box, with final values of
$\sigma_8$ in the range 0.4 -- 1. For all these simulations,
results are available in the form of both the final output
time and earlier times. To be sure that the results were free
from artefacts of the initial conditions, we used only the last
factor of 2 expansion, since at least a factor 3 expansion
is required in order for initial transients to die down
(Baugh et al. 1995). This gave a set of 48 determinations
of $f_{\ss NL}$ to be fitted, from 18 distinct simulations.

\beginfigure{1}
\epsfxsize=8.4cm
\epsfbox[50 415 462 785]{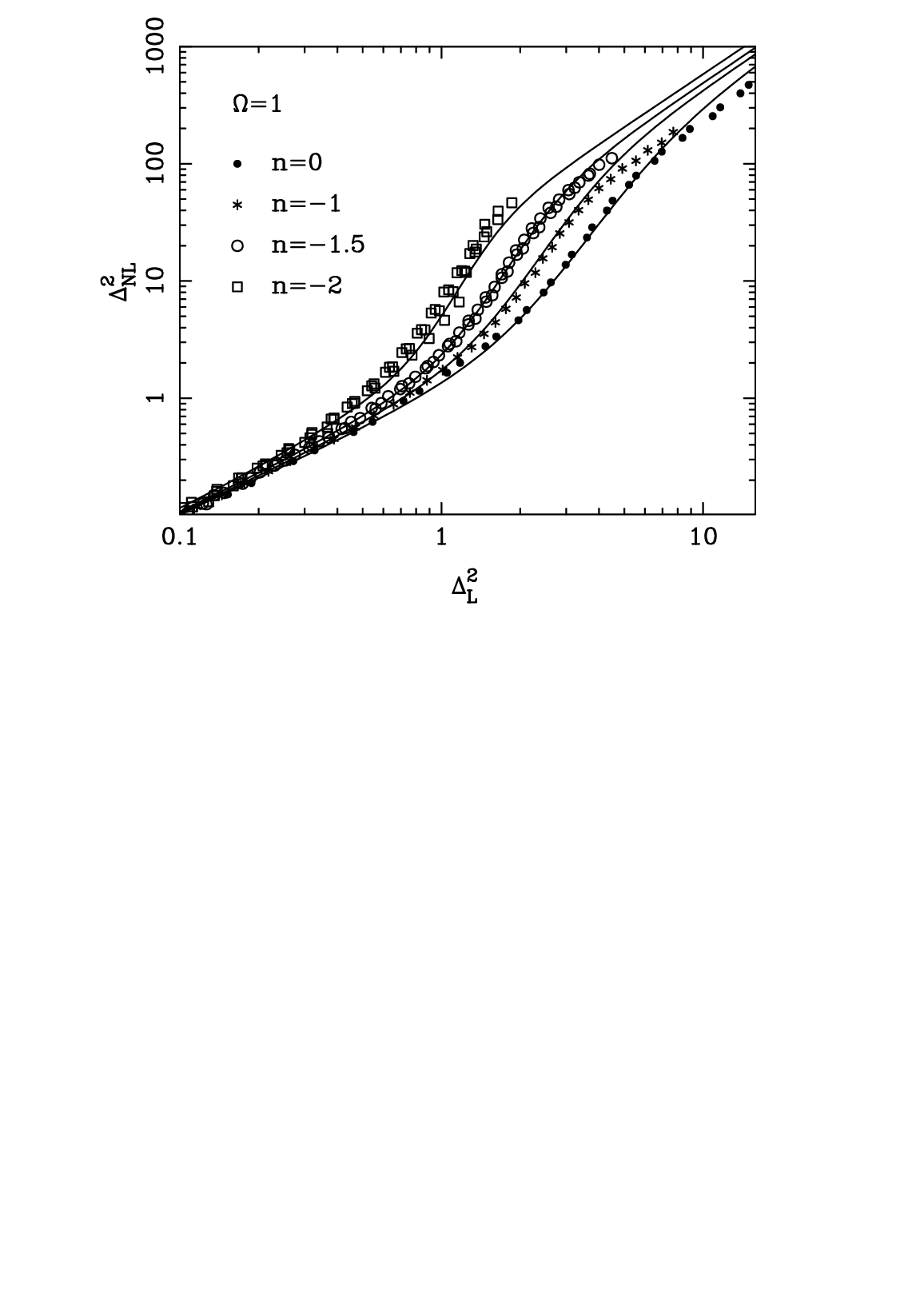}
\caption{%
{\bf Figure 1.}
The generalized HKLM function relating nonlinear power to
linear power, for an $\Omega=1$ Einstein-de Sitter universe
and power-law spectra with $n=0, -1, -1.5, -2$. 
The solid lines show the fitting formula of Section 3.3.
For spectra with $n\gs -1$, the function shows very
little spectral dependence. For flatter spectra, the
nonlinear power at a given linear power is higher;
in particular the slope of the quasilinear portion
around $\Delta^2_{\ss NL}\sim 10$ increases as
$n$ decreases, from approximately $f_{\ss NL}(x)\propto x^3$ for $n=0$
to $f_{\ss NL}(x)\propto x^4$ for $n=-2$.
}
\endfigure

The next few Figures give a selection of results.
Fig. 1 shows $f_{\ss NL}$ for a variety of power-law spectra
with $\Omega=1$, whereas Figs 2 \& 3 shows how these change with
decreasing $\Omega$, both without and with a cosmological constant. Figs 2 \& 3
illustrate the point made by PD, that $f_{\ss NL}$ steepens as
we go to lower-density models. Fig. 1 shows that JMW
were correct in claiming that there was also a significant dependence
on spectral index, particularly for spectra with $n\ls-1$.
Such flat spectra have a larger $f_{\ss NL}$, but
$f_{\ss NL}$ also appears to steepen for more negative $n$,
whereas JMW suggested a fit in which the slope of the
quasilinear portion was independent of $n$; in retrospect,
this steepening can also be seen in their data.
Interestingly, however, the spectrum dependence
is less extreme for low-density models, as may be seen in Fig. 2:
for low densities, the quasilinear portion of $f_{\ss NL}$ has
a similar slope for both $n=0$ and $n=-1.5$.
This is an important hint for a way of achieving an improved
fit to the results. What seems to be happening is that the
main effect of changing the density is to alter the amplitude
of the virialized $f_{\ss NL}(x)\propto x^{3/2}$ asymptote.
JMW suggested that this amplitude was also a function of
spectrum, being larger for $n=-2$ than for $n=0$ by about
a factor 2. This could account nicely for the different degrees
of density-dependent steepening: for $n\simeq -2$, the
asymptote is at a sufficiently  high level that raising it still further
by lowering the density has a small effect on the function.
Conversely, for $n\simeq 0$, the nonlinear function saturates
quite early (for $\Omega=1$) owing to the low level of
the asymptote. We now try to find a fitting formula
to see how well this insight works quantitatively.

\beginfigure{2}
\epsfxsize=8.4cm
\epsfbox[50 15 462 785]{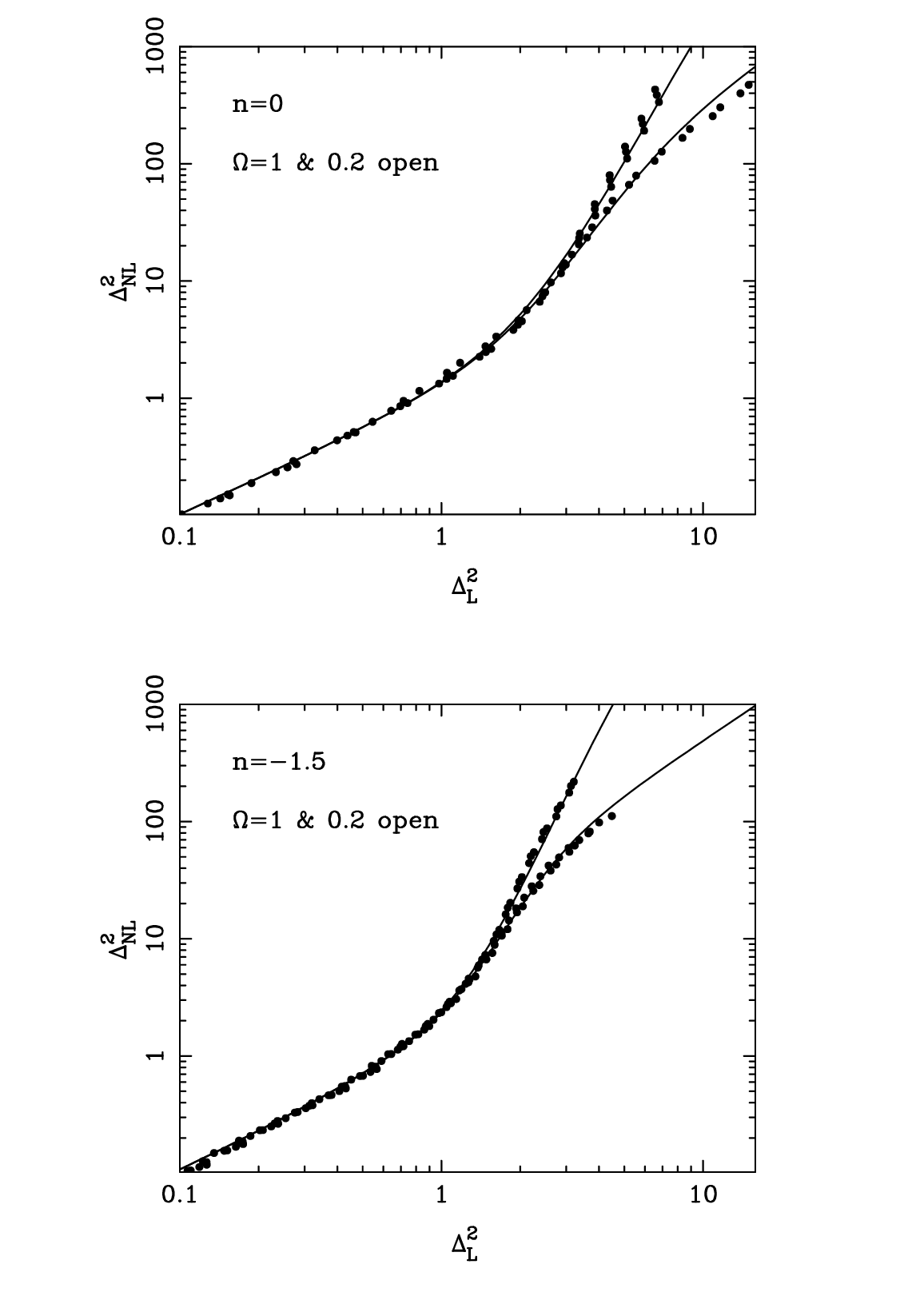}
\caption{%
{\bf Figure 2.}
The generalized HKLM function relating nonlinear power to
linear power, for $\Omega_{\rm m}=1$ \& 0.2 and zero vacuum energy.
(a) an $n=0$ spectrum; (b) $n=-1.5$.
The solid lines show the fitting formula of Section 3.3.
As $\Omega$ decreases, the nonlinear power increases
and the quasilinear portion steepens. However, 
these changes are largely associated
with the increase of $f_{\ss NL}$ in the virialized regime with
$\Delta^2_{\ss NL}\gs 100$.
}
\endfigure

\beginfigure{3}
\epsfxsize=8.4cm
\epsfbox[50 15 462 785]{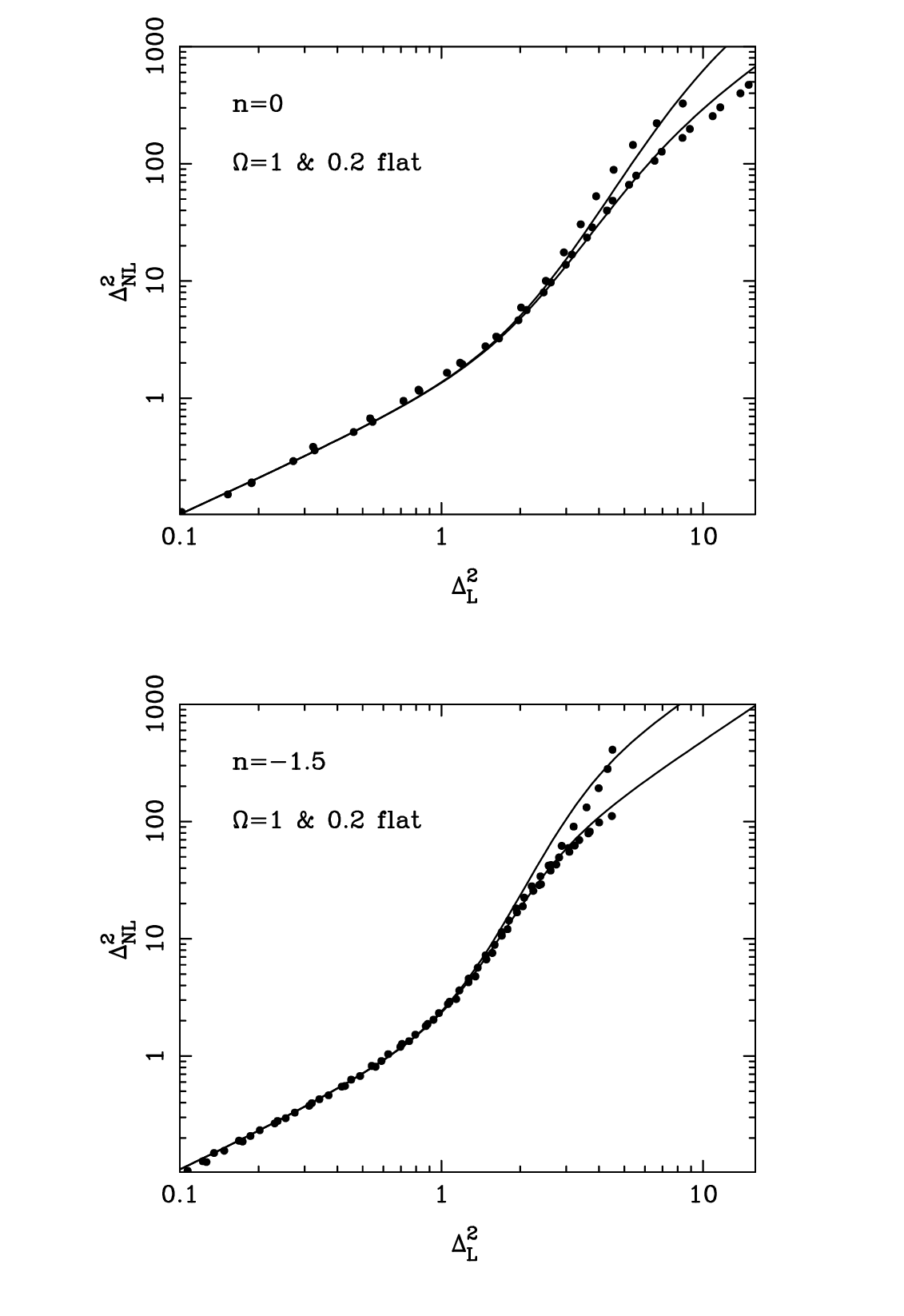}
\caption{%
{\bf Figure 3.}
As for Fig. 2, but for spatially flat models with
$\Omega_{\rm v}+\Omega_{\rm m}=1$.
The generalized HKLM function relating nonlinear power to
linear power, 
for $\Omega_{\rm m}=1$ \& 0.2.
(a) an $n=0$ spectrum; (b) $n=-1.5$.
The solid lines show the fitting formula of Section 3.3.
The $\Omega$ dependence is weaker than for open models,
consistent with the idea that all that matters is the
growth-suppression factor $g(\Omega)$.
}
\endfigure

\ssec{Fitting formula}

The fitting formula is the same form as used by PD:
$$
f_{\ss NL}(x) =x \; \left[{ 1+B\beta x +[A x]^{\alpha\beta} \over
1 + ([A x]^\alpha g^3(\Omega)/[V x^{1/2}])^\beta}\right]^{1/\beta}.
$$
This contains 5 free parameters, each of which is potentially spectrum
dependent. $B$ describes a second-order deviation from linear
growth; $A$ and $\alpha$ parameterise the power-law which
dominates the function in the quasilinear regime; $V$ is
the virialization parameter which gives the amplitude of the
$f_{\ss NL}(x) \propto x^{3/2}$ asymptote; $\beta$ softens
the transition between these regimes.

We proceeded by fitting the $\Omega=1$ results for individual power-law spectra,
and looking at the trends of the parameters with $n$. This
suggested a functional form which became progressively more nonlinear as
$n$ approached $-3$ (as with the power of $(1+n/3)$ proposed by JMW).
The next step was to fit what is now a 10-parameter model
to the data over the range  $-2<n<0$, and this was
achieved satisfactorily, with an overall rms accuracy of about 12 per cent
in $f_{\ss NL}(x)$ over the range
$0.1 < \Delta^2_{\ss NL} <10^{2.5}$. Adding data with $\Omega\ne 1$ required
very little alteration to the fit, as hoped from the discussion in the previous
Section: the $g^3$ term in the fitting formula seems to
be all that is required to incorporate different cosmological models.

\beginfigure{4}
\epsfxsize=8.4cm
\epsfbox[50 15 462 785]{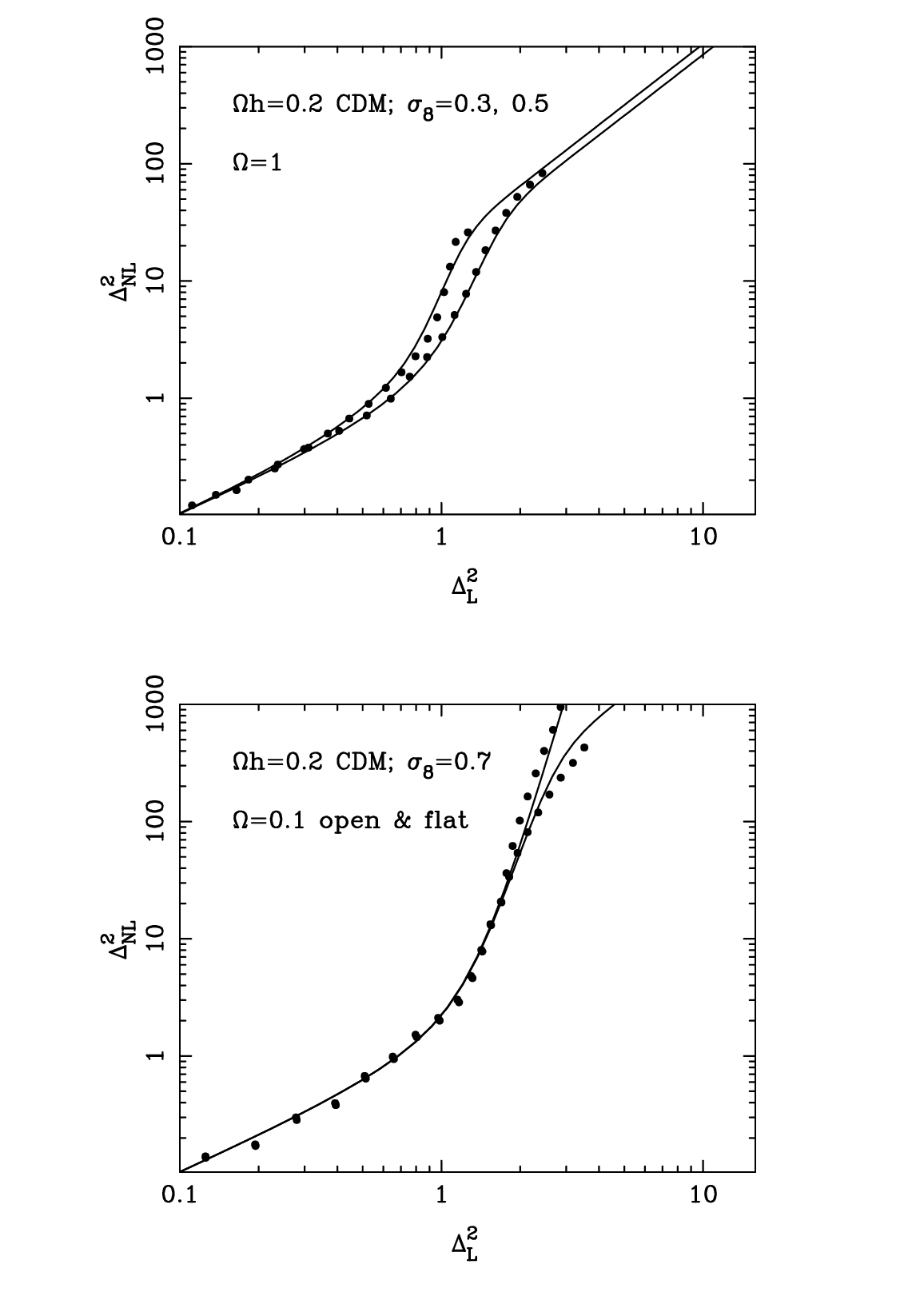}
\caption{%
{\bf Figure 4.}
As for Fig. 2, but for standard CDM spectra.
The generalized HKLM function relating nonlinear power to
linear power, for 
(a) for an $\Omega=1$ Einstein-de Sitter universe; $\Omega h=0.2$
and $\sigma_8=0.3$ \& 0.5.
(b) for an $\Omega_m=0.1$ universe with
$\Omega_v=0$ \& 0.9; $\Omega h=0.2$, $\sigma_8=0.7$.
The solid lines show the fitting formula of Section 3.3.
Note that the CDM curves are steeper even than the
$n=-2$ power-law results, suggesting that the
small-scale CDM behaviour is characteristic of the
tangent spectral index there, which can be as negative
as $n=-2.5$ for the models studied here.
Figure 4b shows clearly the extra small-scale power produced
in the case of an open model by comparison with a
a flat model of the same $\Omega_m$.
}
\endfigure

The final step was to incorporate CDM results, which  are
important for two reasons.
The CDM spectrum is the most important example of a spectrum
which curves slowly so that the effective power-law index
$n_{\rm eff}\equiv d\ln P / d \ln k$ varies with
scale. It is also a useful case for the present investigation,
since it contains very little large-scale power, but has
an effective $n$ that tends to $-3$ on small scales. Pure
power-law spectra with $n<-2$ are very hard to simulate:
being so flat, they tend to saturate the fundamental mode
before enough small-scale nonlinear evolution has occurred
to erase the initial conditions. 
JMW suggested that the nonlinear behaviour of CDM
models could be modeled via the power-law model
corresponding to $n_{\rm eff}$ at the nonlinear scale.
However, it seemed to us that the whole philosophy of the
HKLM method is that the nonlinear power at $k_{\ss NL}$
derives from the linear power at the smaller $k_{\ss L}$.
One would therefore expect that the appropriate
treatment for CDM models would be to use a
different $f_{\ss NL}$ at each $k_{\ss L}$, according
to the tangent spectral index at that point.
This is an important assumption, because it means that
the small-scale power in CDM models should be representative
of the $n<-2$ spectra which are otherwise so hard to treat.
Without this assumption, we found our CDM results hard
to fit: they reach nonlinear powers on the smallest scales greater
than would be expected from even the $n=-2$ power-law
fits (see Fig. 4), but the nonlinear response at larger
scales is less extreme, as would be expected if the
effective $n$ was larger. However, although the trend of
nonlinear response with scale is as expected, the CDM results
generally lay below those predicted from the power-law
fits using $n_{\rm eff}(k_{\ss L})$. This is not unreasonable:
if we compare a CDM spectrum with its tangent power-law spectrum,
the power-law spectrum has a greater amount of power integrated
from $k=0$ to k=$k_{\ss L}$. A practical means for accounting for this
difference is to conjecture that the nonlinear behaviour of
the CDM spectrum will be characteristic of its tangent index
at some slightly smaller scale, and a shift of a factor 2 in $k$
gives outstandingly good results:
$$
n_{\ss L}(k_{\ss L})\equiv {d\ln P \over d \ln k}(k=k_{\ss L}/2).
$$ 
The exact size of the shift is not critical, and we have made
no attempt to treat it as an additional parameter to be optimized.
Very good predictions of nonlinear CDM spectra are achieved even
simply using the unshifted tangent spectral index.
A shift in this sense is likely to be required for any convex
spectrum, where $n_{\rm eff}$ decreases as $k$ increases.
Since this prescription clearly also works for pure power-law
spectra, we propose this as a general method for dealing
with any smoothly-curving convex spectra that are hierarchical
in the sense that $\Delta^2$ increases as $k$ increases.
The obvious exceptions are therefore spectra with a
small-scale cutoff, such as hot or warm
dark matter, and these will require separate treatment.

We therefore performed a global fit to the power-law plus CDM
data with this assumption, and obtained an excellent fit
over all cosmological models, with an rms accuracy of about 14 per cent
in $f_{\ss NL}(x)$. This a demanding level of agreement, since
$f_{\ss NL}(x)$ is so steep; the scatter in the transverse direction,
which governs the accuracy of power-spectrum reconstruction, 
is only about 7 per cent. As discussed above,  errors in $f_{\ss NL}$
partially normalized themselves away when predicting nonlinear
power, and this figure of 7 per cent is also the approximate
rms accuracy with which this method here will predict $\Delta^2_{\ss NL}$
for a given linear spectrum.
The best-fit parameters are
$$
A=0.482\,(1+n/3)^{-0.947}
$$
$$
B=0.226\,(1+n/3)^{-1.778}
$$
$$
\alpha=3.310\,(1+n/3)^{-0.244}
$$
$$
\beta=0.862\,(1+n/3)^{-0.287}
$$
$$
V=11.55\,(1+n/3)^{-0.423}.
$$

Note once again that the cosmological model does not enter anywhere
in these parameters. It is present in the fitting formula
only through the growth factor $g$, which governs the
amplitude of the virialized portion of the spectrum.
This says that all the quasilinear features of the
power spectrum are independent of the cosmological
model, and only know about the overall level of power.
This is not surprising to the extent that quasilinear
evolution is well described by the
Zeldovich approximation, in which the final positions of
particles are obtained by extrapolating their initial
displacements by some universal time-dependent factor.
All information on the cosmological model is hidden in
this extrapolation factor, and therefore the model
should have no effect if we scale to displacements of
the same size. The power spectrum in the Zeldovich
approximation has been calculated analytically by
Taylor (1993) and by Schneider \& Bartelmann (1995), and it would be of interest
to compare their results with ours.

\sec{DISCUSSION}

We have investigated in detail the scaling formula of
HKLM for the evolution of clustering statistics in cosmology. Although not
completely spectrum-independent, their approach
can be made to give a good fit for a variety of
spectra, provided one uses a simple dependence on the
tangent slope of the linear power spectrum.
Because of the need for a spectrum-dependent correction,
we have not provided a fitting formula for $\bar\xi$, nor
for the inverse nonlinear function. 
Inverting observed nonlinear data is now in any case
an iterative process, and $\bar\xi$ is a statistic of less
practical interest than the power spectrum. It is probably
best to proceed numerically with the forward nonlinear
function as a starting point.

\beginfigure{5}
\epsfxsize=8.4cm
\epsfbox[50 15 462 785]{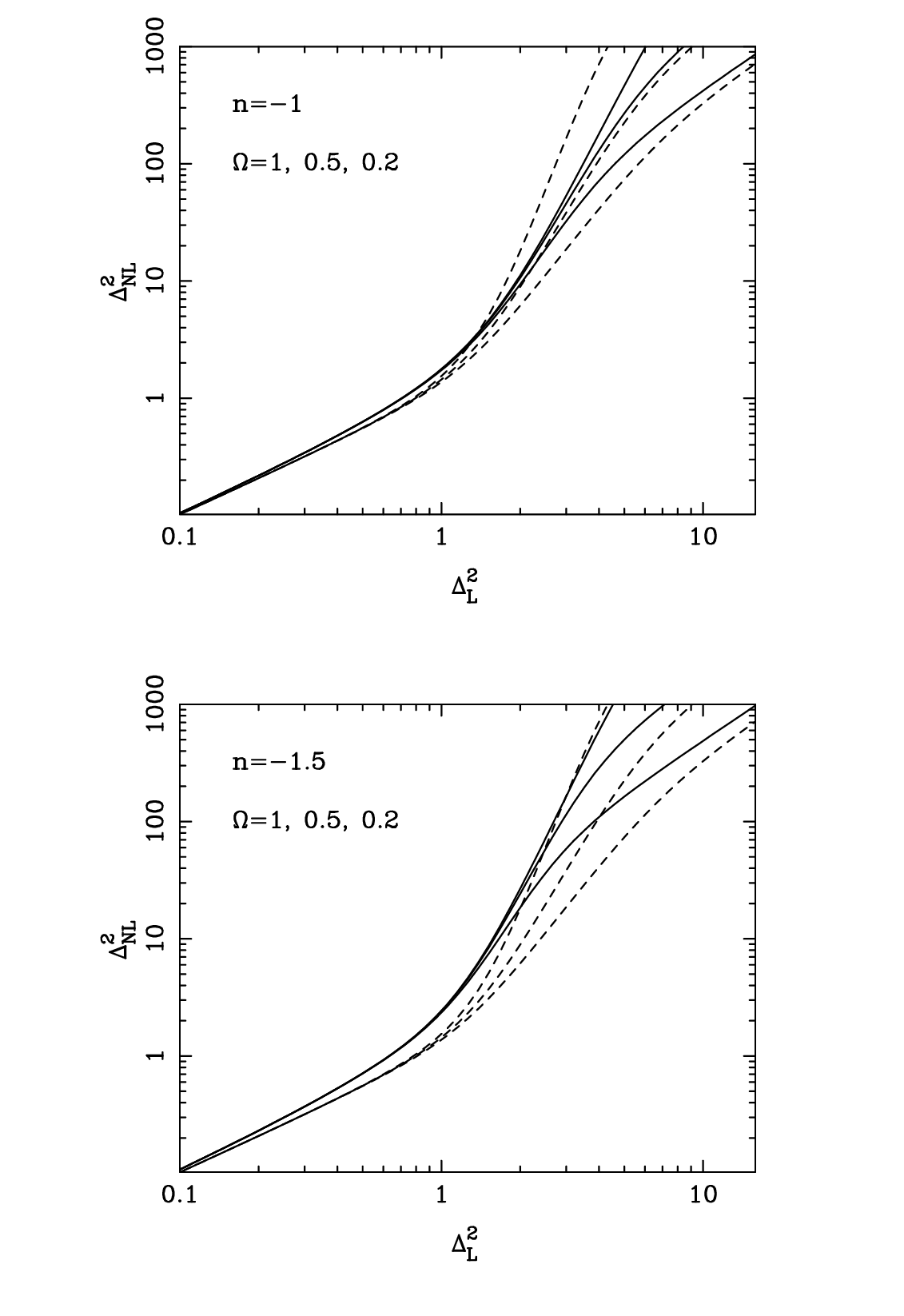}
\caption{%
{\bf Figure 5.}
The present fitting formula for the nonlinear function
(solid lines) compared with the spectrum-independent
form suggested by PD (dashed lines), for the cases
$n=-1$ \& $-1.5$ and open models with $\Omega_m=1,0.5,0.2$.
The PD formula was an approximation to the average effect
of these different spectra, but the detailed agreement is
often poor, particularly at
low $\Omega$ for flatter spectra.
However, these deviations are small
in the regime of the data used by PD ($\Delta^2_{\ss NL}\ls 3$).
}
\endfigure

The main feature of the nonlinear HKLM function in 
this work is a power-law on intermediate scales
which is rather steep, $f_{\ss NL}(x)\propto x^{1+\alpha}$
with $\alpha\simeq 3.5$ -- 4.5,
and it is a challenge to understand this result.
Padmanabhan (1996) has given arguments to suggest that
the intermediate slope should be 
$f_{\ss NL}(x)\propto x^3$, but it seems that
this is not the true value. For spectra
with $n\simeq 0$, such an index works well enough,
but this appears to be an artefact of the relatively
rapid onset of the virialized regime. For lower
densities or very negative $n$, the virialized
regime occurs at larger powers, so that the
steeper intermediate behaviour of $f_{\ss NL}(x)$
is seen more clearly.
This steep quasilinear function has an interesting
implication for the scaling of small-scale power
spectra and their evolution with time.
For $\Delta^2 \gg 1$, we have
$$
\eqalign{
k_{\ss NL} &\simeq [\Delta^2_{\ss NL}]^{1/3} k_{\ss L} \cr
\Delta^2_{\ss NL} &\propto [D^2(a)\, \Delta^2_{\ss L}]^{1+\alpha}, \cr
}
$$
where $D(a)$ is the linear growth law for density perturbations.
For a power-law linear spectrum, this predicts a quasilinear
power law
$$
\Delta^2_{\ss NL}\propto D^{(6-2\beta)(1+\alpha)/3}\, k_{\ss NL}^\beta,
$$
where the nonlinear power-law index depends as follows on
the slope of the linear spectrum:
$$
\beta =  { 3(3+n)(1+\alpha)\over 3+ (3+n)(1+\alpha)}.
$$
For the observed index of $\beta\simeq 1.8$, this would
require $n\simeq -2.2$, very different from the 
$n=0$ that would give $\beta=1.8$ in the virialized
regime. However, especially for low density models,
the virialized regime is only reached on very small
scales and the observed clustering data are dominated
by quasilinear effects. It is then interesting to note
that the predicted evolution is faster than the
linear $\Delta^2\propto D^2(a)$; this may be of relevance
in understanding the weak angular clustering of faint galaxies
(e.g. Efstathiou et al. 1991; Roche et al. 1993).

These results should be of practical use
for a variety of cosmological investigations.
The most obvious case is the theme pursued in PD:
linearizing observed clustering data in an attempt to
infer the underlying linear power spectrum. The
fitting formula used by PD was restricted by the
assumption that $f_{\ss NL}$ was spectrum independent,
and the results given here are of considerably greater
accuracy, particularly for $n\ls -1$. How much this
matters depends on the application of interest;
because $f_{\ss NL}$ is so steep, it is possible to
predict $\Delta^2_{\ss NL}$ rather badly and yet be
able to infer $\Delta^2_{\ss L}$ with tolerable
accuracy. This is illustrated in Fig. 5, which
compares the present fitting formula with the one
given in PD. There are deviations of up to a power
of ten in $\Delta^2_{\ss NL}$ in the case of great nonlinearity and low
density. However, in the regime of the data actually 
used by PD ($\Delta^2_{\ss NL}\ls 3$), the errors
in the deduced $\Delta^2_{\ss L}$ from a given
nonlinear power
are at most around 20 per cent; the work reported here thus does not imply any
serious revision of the conclusions reached in PD.
Nevertheless, there remains the challenge of 
understanding the highly nonlinear portion
of the the power spectrum, and the improved fitting
formula from this paper should be of use in attempts
to interpret the clustering data in this regime.

\secno{ACKNOWLEDGEMENTS}%
SJD was supported by a SERC/PPARC
research studentship during part of this work.
We salute Hugh Couchman for his tremendous contribution to
cosmology in making his AP3M code freely available.
We thank Bhuvnesh Jain for helpful correspondence
on this subject.

\secno{REFERENCES}%
\beginrefs
\ref Baugh C.M., Efstathiou G., 1994, \mn, 270, 183
\ref Baugh C.M.,  Gazta\~naga E., Efstathiou G., 1995, \mn, 274, 1049
\ref Carroll S.M., Press W.H., Turner E.L., 1992, \annrev, 30, 499
\ref Couchman H.M.P., 1991, \apj, 368, L23
\ref Efstathiou G., Bernstein G., Katz N., Tyson T., Guhathakurta P., 1991, \apj, 380, 47
\ref Hamilton A.J.S., Kumar P., Lu E.,  Matthews A., 1991, \apj, 374, L1 (HKLM)
\ref Jain B., Mo H.J., White S.D.M., 1995, \mn\ 276, L25 (JMW)
\ref Lahav O., Lilje P.B., Primack J.R., Rees, M.J., 1991, \mn, 251, 128
\ref Padmanabhan T., 1996, \mn, 278, L29
\ref Peacock J.A., Nicholson D., 1991, \mn, 253, 307
\ref Peacock J.A., Dodds S.J., 1994, \mn, 267, 1020 (PD)
\ref Peebles P.J.E., 1980, The Large-Scale Structure of the Universe.  Princeton Univ. Press, Princeton, NJ
\ref Roche N., Shanks T., Metcalfe N., Fong R., 1993, \mn, 263, 360
\ref Schneider P., Bartelmann M., 1995, \mn, 273, 475
\ref Taylor A.N., 1993, in Proc. Cosmic Velocity Fields, 9$^{th}$ IAU
Conf, eds F. Bouchet, M. Lachi\`eze-Rey, Editions Fronti\`{e}res, Gif-sur-Yvette,
p585

\endrefs

\bye